\documentclass[conference]{IEEEtran}
\IEEEoverridecommandlockouts
\usepackage{cite}
\usepackage{amsmath,amssymb,amsfonts}
\usepackage{algorithmic}
\usepackage{graphicx}
\usepackage{textcomp}
\usepackage{xcolor}
\usepackage{tabularray}
\usepackage{enumitem}
\usepackage{subcaption}
\usepackage{float}
\usepackage[labelsep=space]{caption} 
\usepackage[export]{adjustbox}
\graphicspath{{pic/}}
\usepackage{multirow}
\usepackage[section]{placeins}
\usepackage{booktabs}
\usepackage{threeparttable}

\usepackage{color, soul, framed}
\usepackage{multicol} 
\usepackage{pifont} 
\usepackage{caption}
\usepackage{stfloats}

\def\BibTeX{{\rm B\kern-.05em{\sc i\kern-.025em b}\kern-.08em
    T\kern-.1667em\lower.7ex\hbox{E}\kern-.125emX}}
    
\begin{document}

\title{A Multi-Grid Implicit Neural Representation for Multi-View Videos\\

\thanks{This work was partly supported by the NSFC62431015, Science and Technology Commission of Shanghai Municipality No.24511106200, the Fundamental Research Funds for the Central Universities, Shanghai Key Laboratory of Digital Media Processing and Transmission under Grant 22DZ2229005, 111 project BP0719010, Okawa Research Grant and SJTU-Waseda Seed Research Fund. (Corresponding Author: Zhengxue Cheng.)
}

\vspace{-13pt}

\author{Qingyue Ling$^{1}$, Zhengxue Cheng$^{1}$, Donghui Feng$^{1}$, Shen Wang$^{1}$, Chen Zhu$^{2}$, Guo Lu$^{1}$, \\Heming Sun$^{3}$, Jiro Katto$^{4}$, Li Song$^{1}$ \\
$^{1}$Institute of Image Communication and Network Engineering, Shanghai Jiao Tong University. \\
$^{2}$ Nvidia. \\
$^{3}$ Faculty of Engineering, Yokohama National University. \\
$^{4}$ Graduate School of Fundamental Science and Engineering, Waseda University. 
\vspace{-13pt}
}
}

\maketitle

\begin{abstract}
Multi-view videos are becoming widely used in different fields, but their high resolution and multi-camera shooting raise significant challenges for storage and transmission. In this paper, we propose MV-MGINR, a multi-grid implicit neural representation for multi-view videos. It combines a time-indexed grid, a view-indexed grid and an integrated time and view grid. The first two grids capture common representative contents across each view and time axis respectively, and the latter one captures local details under specific view and time. Then, a synthesis net is used to upsample the multi-grid latents and generate reconstructed frames. Additionally, a motion-aware loss is introduced to enhance the reconstruction quality of moving regions. The proposed framework effectively integrates the common and local features of multi-view videos, ultimately achieving high-quality reconstruction. Compared with MPEG immersive video test model TMIV, MV-MGINR achieves bitrate savings of 72.3\% while maintaining the same PSNR.
\end{abstract}

\begin{IEEEkeywords}
Multi-View Videos, Implicit Neural Representation, Multi-Grid Latents, Motion-Aware Loss
\end{IEEEkeywords}


\section{Introduction}
Driven by advancements in computer vision and immersive media, multi-view videos are widely used in the fields of virtual reality, augmented reality and 3D reconstruction \cite{ref1}. However, due to the characteristics of high-resolution and multi-camera shooting, the large data volume of multi-view videos leads to higher storage and transmission costs. Traditional coding standards\cite{ref2,ref3} for multi-view videos mainly eliminate the inter-view redundancy by calculating pixel-domain residuals to achieve disparity compensation. However, the reconstruction quality highly relies on high-precision depth maps and camera calibration parameters. The high computational complexity of these methods also leads to low coding efficiency. 

In recent years, implicit neural representation (INR) shows great potential in various vision tasks \cite{ref4,ref5,ref6,ref7,ref8,ref9,ref10,ref11,ref12}. It interprets the signal as a coordinate-to-value mapping by encoding them into a compact parameterized representation using a neural network. Then the network is trained to approximate this mapping. For instance, COIN \cite{ref4} models a signal by overfitting a multilayer perceptron (MLP) that maps pixel coordinates to RGB values. NeRV \cite{ref7} models video signals along the temporal dimension and uses a convolutional structure to simplify the model parameters. However, applying INR to multi-view video compression still face a few challenges. The large data volume of multi-view videos significantly increases the computational complexity. The neural network needs to store and optimize weights for each spatio-temporal position, leading to low processing efficiency. To address this, MV-IERV\cite{ref13} takes time and view index of multi-view videos as coordinate inputs and utilizes high-quality reconstructed frames from explicit codec to achieve inter-view compensation. However, its reconstruction quality of moving regions still need to be improved, which is also a challenge for other INR-based methods. Therefore, it requires to propose an efficient and effective way for multi-view video compression.

\begin{figure}[t]
    \centering
    \includegraphics[width=\linewidth]{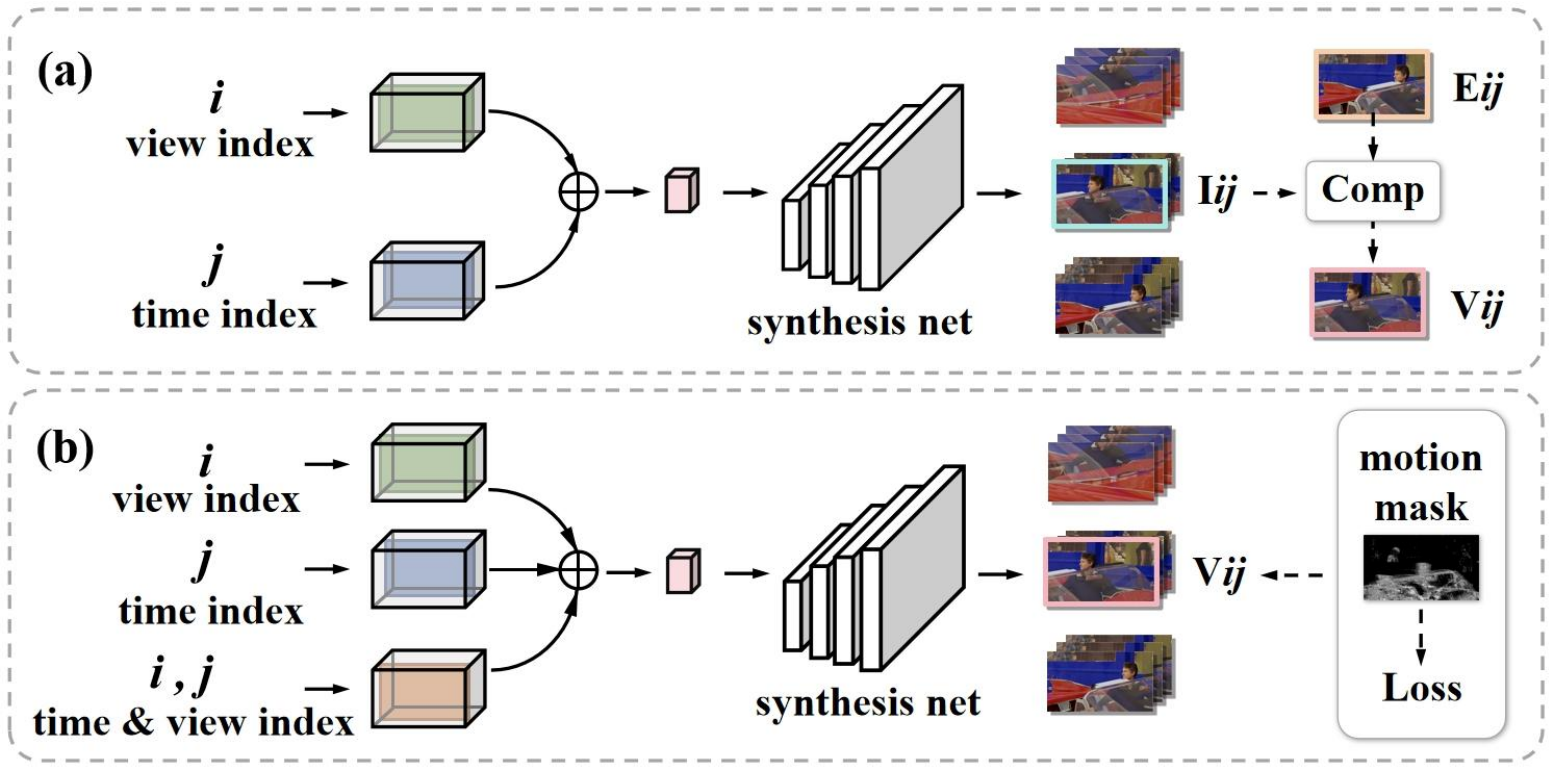}
    \caption{(a) MV-IERV\cite{ref13}. A time-indexed grid and a view-indexed grid are used to obtain the input latents for synthesis net. Besides, explicit reconstruction is used for inter-view compensation (b) The proposed MV-MGINR. Another integrated time and view grid is introduced to improve the reconstruction quality and motion mask is incorporated into training loss.}
    \label{fig1}
    \vspace{-18pt}
\end{figure}

To address the above problems, we propose a multi-grid implicit neural representation for multi-view videos, named MV-MGINR, as shown in Fig.~\ref{fig1}. It consisits of a time-indexed grid, a view indexed grid and an integrated time and view grid. The first two grids capture common representative contents across each time and view axis respectively. The latter one captures local details under specific view and time to enhance the reconstruction quality. The obtained multi-grid latents are then processed through a synthesis net to generate final implicit reconstructed frames. Moreover, explicit motion masks are incorporated into training loss to improve the reconstruction quality of moving regions.

We evaluate MV-MGINR on MPEG immersive video (MIV) standard dataset \cite{ref14}. Results show that it outperforms the MIV test model TMIV \cite{ref3} and other learning-based or INR-based codecs \cite{ref7,ref13,ref16,ref17,ref28} in terms of multi-view compression and reconstruction performance.

Our contributions can be summarized as follows:

\begin{figure*}[htbp]
    \centering
    \includegraphics[width=\linewidth]{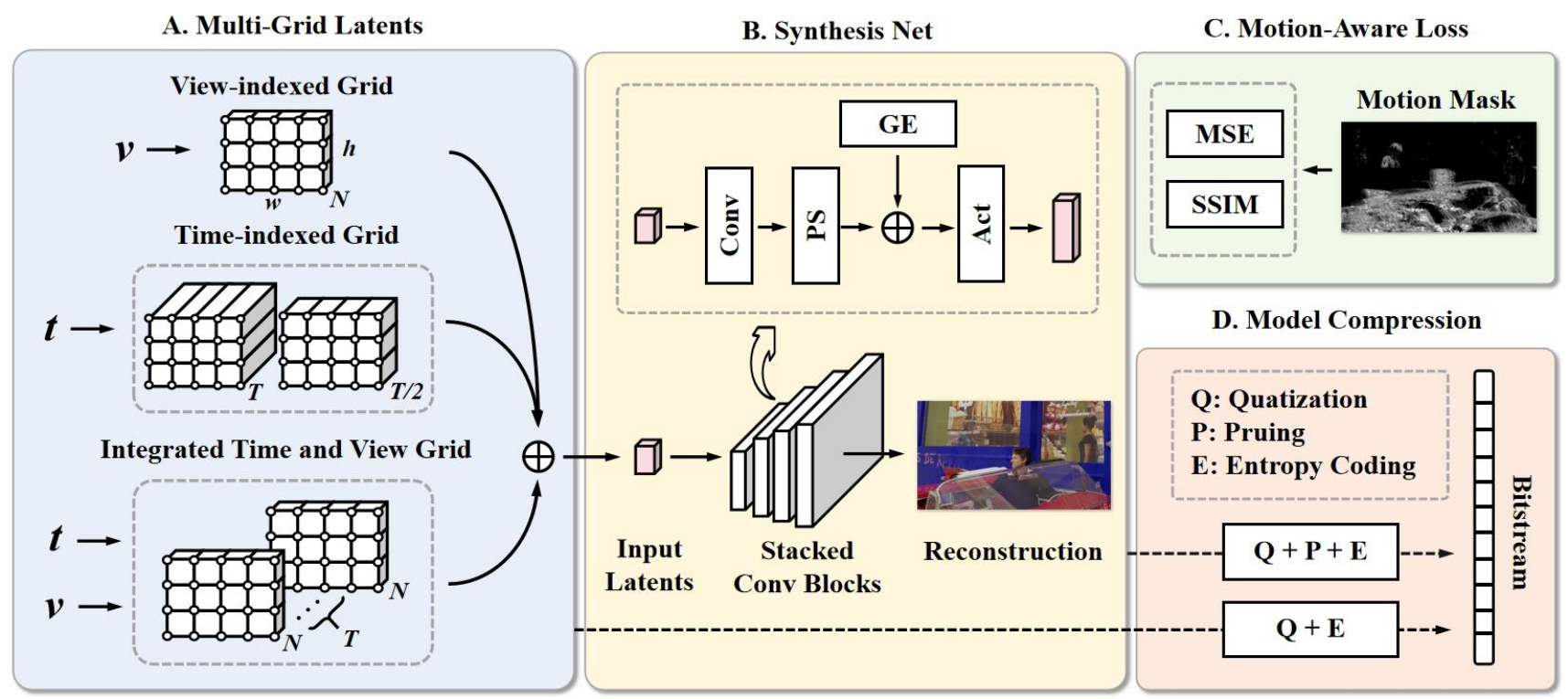}
    \caption{ Overview of MV-MGINR. A. Input latents can be obtained from multi-grid with time and view index. B. The synthesis net then processes the input latents and generates the reconstructed frame. C. By computing optical flow, motion mask is incorporated into training loss. D. Quantization, pruning and entropy coding are used for model compression.}
    \label{fig2}
    \vspace{-15pt}
\end{figure*}

\begin{itemize}
\item  We propose a multi-grid implicit neural representation for multi-view videos. It effectively integrates the common and local features of multi-view videos under different views and time, achieving high-quality reconstruction.

\item  The reconstruction quality of moving regions is improved by introducing a motion-aware loss during training. 

\item  Experimental results show that MV-MGINR achieves bitrate savings of 72.3\% while maintaining the same PSNR compared with TMIV, which indicates the effectiveness of the proposed method.
\end{itemize}


\vspace{-2mm}

\section{Related Work}

Previous multi-view video compression standards, such as MV-HEVC and 3D HEVC\cite{ref27}, primarily reduce inter-view redundancy by using disparity vectors (DV). The Moving Picture Experts Group (MPEG) developed TMIV\cite{ref3}, which divides the source views into basic views and additional views. It then removes inter-view redundancy of additional views and combines them with the basic view data to produce a flat atlas, which is finally compressed by a standard 2D video codec.

INR uses neural networks to encode each signal into a compact parameterized representation. Due to its powerful ability to express complex signals, it is also widely used in video compression\cite{ref4,ref7,ref17,ref19,ref20,ref21,ref24,ref25,ref26,ref37}. For multi-view videos, MV-HiNeRV\cite{ref28} learns different grids for each view and shares the learnt network parameters, effectively exploiting the spatio-temporal and the inter-view redundancy that exists within multi-view videos. MV-IERV\cite{ref13} takes time and view index of multi-view videos as inputs of grids and utilizes high-quality reconstructed frames from explicit codec to achieve inter-view compensation. However, due to the large data volume of multi-view video formats, higher compression efficiency and better reconstruction quality are required, leaving room for further development in the field of multi-view video compression.

\section{Proposed Method}

We propose MV-MGINR, consisting of a multi-grid, a synthesis net and a motion-aware loss. Model compression methods are then adopted to compress the model size. The proposed framework effectively integrates the common and local features of multi-view videos and ultimately achieves high-quality reconstruction.

\vspace{-3pt}
\subsection{Multi-Grid Latents}
\vspace{-3pt}

The problem can be set as follows: given a multi-view video sequence $V$ with $N$ views, each of which contains $T$ frames, with resolution of $H \times W$. As illustrated in Fig.~\ref{fig2}, we propose a multi-grid consisting of a time-indexed grid, a view-indexed grid and an integrated time and view grid.

\textbf{Time-indexed grid.} Inspired by \cite{ref13}, a multi-resolution grid, taking time index as input, is used to extract the temporal features along time dimension. The gird $G_{time}$ and the obtained temporal latents $l_{time}$ can be expressed as follows:

\vspace{-2mm}
\begin{gather}
G_{time}=\left\{ G_{time}^{1},G_{time}^{2} \right\} \\
l_{time}^{1}=G_{time}^{1}\left( \text{Floor}\left( t \right) \right), l_{time}^{2}=G_{time}^{2}\left( \text{Floor}\left(t/2 \right)\right) 
\end{gather}
where $G_{time}^{1}$ and $G_{time}^{2}$ represent time-indexed grids of different resolution. $l_{time}^{1}$ and $l_{time}^{2}$ represent the obtained temporal latents from $G_{time}^{1}$ and $G_{time}^{2}$, respectively. $\text{Floor}\left( \cdot \right) $ is the floor function. $t$ represents the time index. The two temporal latents of different resolution are then concatenated to form the final temporal latents, donated as $l_{time}\in \mathbb{R}^{1\times h\times w\times c_1}$, where $h,w,c_1$ represent the height, width and channel number of the temporal latents from time-indexed grid. 

\textbf{View-indexed grid.} The view-indexed grid $G_{view}$ is used to extract the perspective features and obtain the perspective latents $l_{view}$, which can be described as follows:

\vspace{-4mm}
\begin{gather}
l_{view}=G_{view}\left( v \right)
\end{gather}
where $l_{view}\in \mathbb{R}^{1\times h\times w\times c_1}$ has the same shape of temporal latents $l_{time}$. $v$ represents the view index. 

Since $G_{time}$ and $G_{view}$ are shared by all view indexes and time indexes, the obtained latents $l_{time}$ and $l_{view}$ represent common features across each view and time axis, respectively. Finally, the two latents are added to form latents $l_{tv1}$.

\textbf{Integrated time and view grid.} $G_{time}$ and $G_{view}$ lack the capability to extract features under specific time and view, especially in videos with significant content changes. Therefore, we add an integrated time and view grid that can further extract local features of specific frame. The integrated time and view grid $G_{tv}$ and the obtained latents  $l_{tv2}$ can be expressed as follows:

\vspace{-4mm}
\begin{gather}
l_{tv2}=G_{tv}\left( t,v \right) 
\end{gather}
where $l_{tv2}\in \mathbb{R}^{1\times h\times w\times c_{2}}$. $c_{2}$ represents the channel number of the latents from integrated time and view grid. It's worth noting that $G_{tv}\in \mathbb{R}^{T\times N\times h\times w\times c_2}$, and the size of it will increase significantly due to the multiplicative effect of its dimensions, especially when $T$ and $N$ are large. To avoid too many grid parameters transmitted to the decoder side, which will affect the final bitrate, $c_{2}$ needs to be controlled at a proper value. Finally, $l_{tv1}$ and $l_{tv2}$ are concatenated to obtain the final input latents $l_{input}\in \mathbb{R}^{1\times h\times w\times c}$ for the subsequent synthesis net, where $c=c_{1}+c_{2}$.

\vspace{-4pt}
\subsection{Synthesis Net}
\vspace{-3pt}

As shown in Fig.~\ref{fig2}, the sythesis net consists of $M$ stacked convolutional blocks \cite{ref7}, each of which includes a convolutional layer (Conv), a pixelshuffle layer (PS), and an activation layer (Act). Additionally, the grid embedding structure (GE)\cite{ref13} is incorporated to further improve the representation capability of implicit network. Each convolutional block gradually upsamples the latents size and adjusts dimensions to match the final frame size. A head layer is then connected to the last convolution block to generate the final reconstructed frame.

\vspace{-4pt}
\subsection{Motion-Aware Loss}
\vspace{-3pt}
L1 and SSIM loss are used in the loss function. To further improve the reconstruction quality of moving regions, an explicit motion mask is incorporated into training loss. The final loss function can be denoted as:

\vspace{-3mm}
\begin{equation}
    \begin{split}
        L = &\Big[ \alpha \, \text{L1}(R_{ij}, F_{ij}) + (1 - \alpha) \\
            &\cdot \big(1 - \text{SSIM}(R_{ij}, F_{ij})\big) \Big] \cdot \text{Motion}(V_i)
    \end{split}
\end{equation}
where $R_{ij}$ and $V_{ij}$ represent $j\text{th}$ reconstructed and source frame of the $i\text{th}$ view. $\alpha$ balances the weight of losses. $\text{Motion}\left( \cdot \right)$ represents the explicit motion mask with the range of $\left[ 0,1 \right]$ by calculating opticalflow for each view. The higher value indicates greater motion intensity, enabling the model to pay more attention to moving regions.

\vspace{-2pt}
\subsection{Model Compression and Bitstream}
\vspace{-2pt}

Following \cite{ref7,ref13}, we formulated video compression as model compression. Firstly, global unstructured pruning with adaptive thresholding is used to remove redundant model weights. Next, quantization-aware training\cite{ref29} is adopted for uniform weight quantization. Huffman coding \cite{ref30} then further compresses these values losslessly. The final bitstream transmitted to the decoder-side is composed of parameters from multi-grid and synthesis net.

\vspace{-5pt}
\begin{figure}[tbp]
    \centering 
    \includegraphics[width=0.5\textwidth]{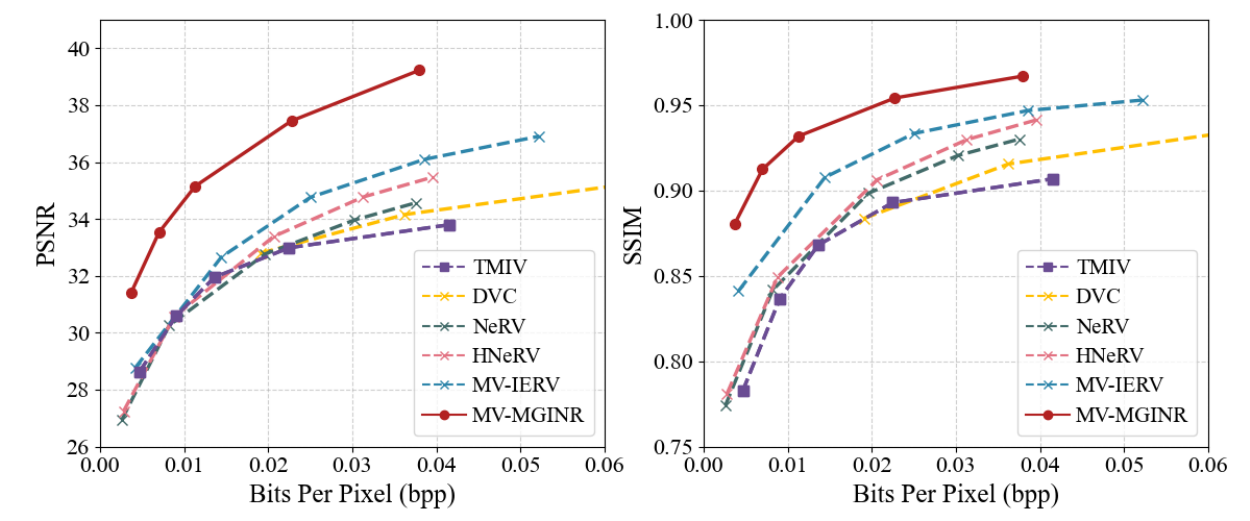} 
    \caption{RD curves measured in PSNR (left) and SSIM (right) of different codecs on MIV dataset.}
    \label{fig3}
    \vspace{-18pt}
\end{figure}

\vspace{-1pt}
\section{Experiment}
\vspace{-2pt}
\subsection{Datasets and Implementation Details}
\vspace{-2pt}
We evaluate our framework on seven videos selected from the MIV standard dataset \cite{ref14}. Each video has a view number $N$ ranging from 9 to 25, an encoded frame number $T$ of 100, and a resolution of $1920\times1080$. In the multi-grid, the height $h$ and width $w$ of the latents are set to 9 and 16. The channel number $c$ of the input latents is specified as $\left\{20,30,40,60,80\right\}$. And the ratio of $c_{1}$ and $c_{2}$ is set to 9:1. In the synthesis net, we use 5 convolutional blocks with upscale factor of $5,3,2,2,2$ respectively. For the loss function, $\alpha$ is set to 0.7 and Farneback opticalflow\cite{ref36} method is adopted to calculate the motion mask for each view.

During the experiments, we trained each test sequences for 300 epochs with batchsize of 2. 8-bit quantization and model pruning with 40\% sparsity are performed to compress model size.  Furthermore, the model is fine-tuned for another 100 epochs to further refine the parameters. PSNR and SSIM\cite{ref32} are used to evaluate the reconstruction quality. We also calculate BD-rate\cite{ref33} to quantitative evaluate the RD performance. All experiments are run with NVDIA RTX4090.

The baseline includes TMIV and several learning-based and INR-based codecs DVC\cite{ref16}, NeRV\cite{ref7}, HNeRV\cite{ref17}, MV-HiNeRV\cite{ref28} and MV-IERV\cite{ref13}. Notably, some of these codecs encode different views independently, so bitrate of all views need to be summed up for total bitrate.

\vspace{-5pt}
\subsection{Quantitative and Qualitative Results}
\vspace{-3pt}

\begin{figure*}[t]
    \centering
    \includegraphics[width=\textwidth]{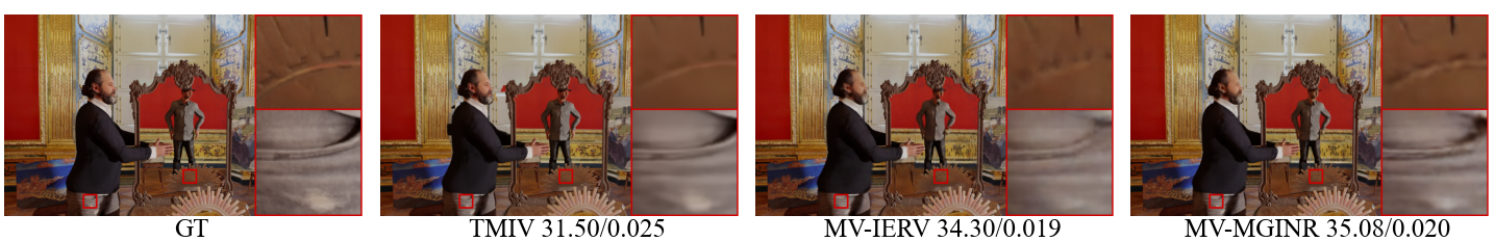} 
    \includegraphics[width=\textwidth]{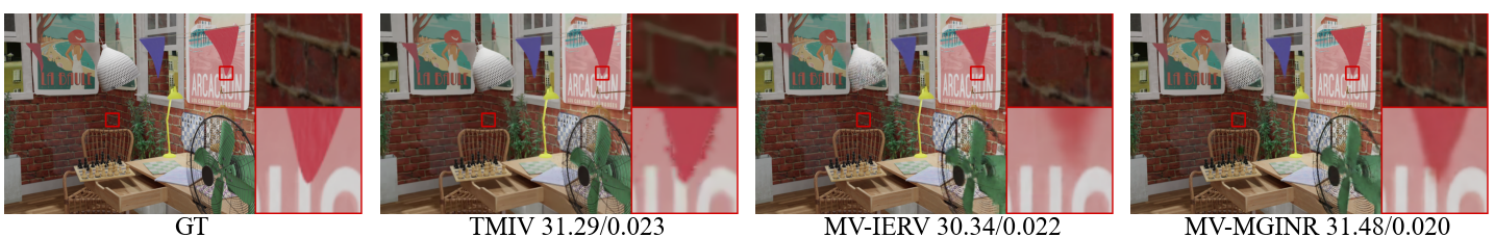}
    \includegraphics[width=\textwidth]{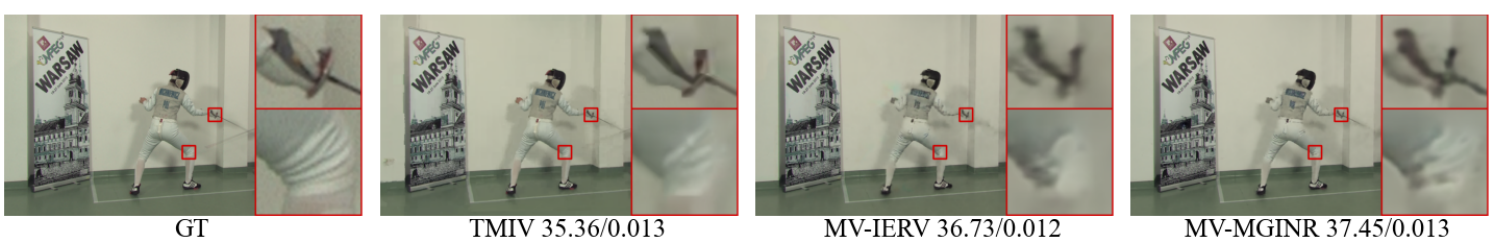}
    \vspace{-10pt}
    \caption{Visualization comparison of Ground Truth, reconstruction of TMIV\cite{ref3}, MV-IERV\cite{ref13} and MV-MGINR. (PSNR/bpp)}
    \label{fig4}
    \vspace{-9pt}
\end{figure*}

After conducting experiments on seven MIV sequences, we calculate the average results to plot the RD curves and compute BD-rate, shown in Fig.~\ref{fig3} and Table~\ref{tab1}. The detailed sequence-level BD-rate results are shown in Table~\ref{tab2}. The results show that compared with TMIV, MV-MGINR achieves bitrate savings of 72.3\% and 80.2\% under the same PSNR and SSIM. As shown in Fig.~\ref{fig4}, we also provide visualization comparison of different methods. It can be observed that the proposed method has a significant improvement in reconstruction quality, especially in the moving regions. 

\begin{table}[tbp]
\vspace{-5pt}
\renewcommand\arraystretch{1.2}
\setlength\tabcolsep{7pt}
        \caption{BD-rate (\%) results of different codecs on MIV dataset, with TMIV\cite{ref3} as anchor.}
\vspace{-8pt}
\begin{center}
\begin{tabular}{ c |  c  c }
\toprule[1.1pt]
Codecs & BD-rate (PSNR) & BD-rate (SSIM)  \\
\hline
NeRV~\cite{ref7} & -0.5 & -16.9 \\
HNeRV~\cite{ref17} & -10.4 & -24.3 \\
DVC~\cite{ref16} & -12.2 & -3.4 \\
MV-IERV~\cite{ref13} & -36.5 & -48.2 \\
MV-HiNeRV~\cite{ref28} & -49.3 & - \\
\textbf{MV-MGINR (Ours)} & \textbf{-72.3} & \textbf{-80.2} \\

\bottomrule[1.1pt]
\end{tabular}
\label{tab1}
\end{center}
\vspace{-12pt}
\end{table}

\begin{table}[tbp]
\renewcommand\arraystretch{1.2}
\setlength\tabcolsep{3.3pt}
\caption{Sequence-level BD-rate (\%) of MV-MGINR on MIV dataset, with TMIV\cite{ref3} as anchor.}
\vspace{-8pt}
\begin{center}
\begin{tabular}{ c |  c  c c c c c c}
\toprule[1.1pt]
Sequence & J02 & J03 & J04  & E02  & E03 & W04   & L01 \\
\hline
BD-rate (PSNR) & -43.36 & -68.49 & -27.14 & -97.63 & -96.37 & -53.31 & -62.18\\
BD-rate (SSIM) & -78.50 & -91.85 & -67.51 & -99.85 & -99.15 & -64.87 & -98.81\\
\bottomrule[1.1pt]
\end{tabular}
\label{tab2}
\end{center}
\vspace{-20pt}
\end{table}

In addition, we test the impact of the compression methods on reconstruction quality, as shown in Table~\ref{tab3}. After quantization, pruning and entropy coding, the number of bits can be greatly reduced while maintaining a similar quality of the original model.

\begin{table}[t]
\vspace{-5pt}
\renewcommand\arraystretch{1.2}
\setlength\tabcolsep{10.5pt}
\caption{Reconstruction quality with different compression methods: Quantization (Q), Pruning (P), Entropy Coding (E).}
\vspace{-4pt}
\begin{tabular}{ccc|c|cc}
\toprule[1.1pt]
Q &  P & E &  bpp & PSNR $\uparrow$ & SSIM $\uparrow$\\
\hline
\ding{55} (FP32) & \ding{55} & \ding{55} & 0.0665 & 35.18 & 0.9369\\ 
\ding{51}  (INT8) & \ding{55} & \ding{55} & 0.0166 & 34.84 & 0.9359\\ 
\ding{51}  (INT8) & \ding{51} & \ding{55} & 0.0085 & 33.53 & 0.9125\\ 
\ding{51}  (INT8) & \ding{51} & \ding{51} & 0.0070 & 33.53 & 0.9125 \\
\bottomrule[1.1pt]
\end{tabular}
\label{tab3}
\vspace{-8pt}
\end{table}

\vspace{-5pt}

\begin{table}[htbp]
\renewcommand\arraystretch{1}
\setlength\tabcolsep{4pt}
\caption{Ablation study results.}
\vspace{-10pt}
\begin{center}
\begin{tabular}{ c | c | c  c }
\toprule[1.1pt]

Method & bpp & PSNR $\uparrow$ & SSIM  $\uparrow$ \\
\midrule
Original                                & 0.0166 & \textbf{34.84} & \textbf{0.9359}       \\
w/o time-indexed and view-indexed grids & 0.0167 &  33.87 & 0.9088   \\
w/o integrated time and view grid       & 0.0165 &  33.83 & 0.9320\\
w/o motion mask                         & 0.0166 &  34.67 & 0.9359\\
\midrule
Original                               &  0.0265 & \textbf{36.57} & \textbf{0.9517} \\
w/o time-indexed and view-indexed grids&  0.0266 & 35.43 & 0.9301   \\
w/o integrated time and view grid      &  0.0263 & 35.46  & 0.9473\\
w/o motion mask                        &  0.0265 & 36.29 & 0.9512\\

\bottomrule[1.1pt]
\end{tabular}
\label{tab4}
\end{center}
\vspace{-25pt}
\end{table}

\begin{table}[t]
\vspace{-3pt}
\renewcommand\arraystretch{1.1}
\setlength\tabcolsep{10pt}
\caption{Model Parameter Distribution.}
\vspace{-10pt}
\begin{center}
\begin{tabular}{ c |  c  c c c}
\toprule[1.1pt]
bpp & $G_{time}$ & $G_{view}$  & $G_{tv}$ & synthesis net \\
\hline
 0.0166 & 5.39\% &  0.97\% &  10.71\% &  83.00\% \\
 0.0265 & 4.30\% &  0.77\% &   8.46\% &  86.48\% \\
\bottomrule[1.1pt]
\end{tabular}
\label{tab5}
\end{center}
\vspace{-8pt}
\end{table}

\vspace{-2pt}
\subsection{Ablation Studies.}
\vspace{-4pt}
To evaluate the contribution of each key component in the proposed method, ablation studies are conducted. We isolate and quantify the effects of multi-grid and the motion-aware loss. Specifically, the multi-grid ablation study comprises two sub-experiments: (i) model without time-indexed and view-indexed grids, (ii) model without the integrated time and view grid. Crucially, we carefully maintain the degraded model with the same size of the original model by adjusting the remaining grid channels, ensuring fair comparison of reconstruction quality under the same bitstream size. The results of the ablation studies are presented in {Table}~\ref{tab4}. Additionally, to better understand the contribution of each model component, we provide the model parameter distribution in {Table}~\ref{tab5}.

\textbf{Effect of multi-grid.} Compared to only using either time-indexed grid and view-indexed grid or integrated time and view grid, the multi-grid method achieves PSNR gains of 1.01dB and 0.97dB at around 0.0166bpp and 1.11dB and 1.14dB at around 0.0265bpp. It attributes to that the multi-grid has better abilities to capture common and local features of multi-view videos, thus achieving high-quality reconstruction.

\textbf{Effect of motion mask.} Compared to only using L1 and SSIM loss as loss function, the motion-aware loss can pay more attention to the reconstruction performance of moving objects, which overcomes the limitations of other INR-based methods in expressing moving regions to a certain extent.

\begin{table}[tbp]
\vspace{-3pt}
\renewcommand\arraystretch{1.1}
\setlength\tabcolsep{10pt}
\caption{Model Complexity Analysis}
\vspace{-10pt}
\begin{center}
\begin{tabular}{ c |  c  c }
\toprule[1.1pt]
Codec & Encoding FPS $\uparrow$ & Decoding FPS $\uparrow$ \\
\hline
TMIV\cite{ref3} & 0.22 & 2 \\
MV-MGINR (Ours) & 0.03 & 82  \\
\bottomrule[1.1pt]
\end{tabular}
\label{tab6}
\end{center}
\vspace{-22pt}
\end{table}

\vspace{-5pt}
\subsection{Complexity Analysis}
\vspace{-3pt}
As shown in Table~\ref{tab6}, similar to the existing INR-based methods, the proposed method has a fast decoding speed, achieving about 82 FPS compared with 2 FPS for TMIV on the MIV dataset. However, since INR-based methods need online
training, the encoding speed is relatively slow. This is also the future direction for improvement.

\vspace{-4pt}
\section{Conclusion}
\vspace{-4pt}
We propose a multi-grid implicit neural representation for multi-view videos. It consists of a time-indexed grid, a view-indexed grid and an integrated time and view grid. The first two grids capture common features across each time and view axis. And the latter one captures local features at specific view and time. The obtained multi-grid latents are then processed through a synthesis net to generate final reconstructed frames. Furthermore, we introduce a motion-aware loss to improve the reconstruction quality of moving regions. Experimental results show that compared to TMIV and other learning-based and INR-based codecs, the proposed method has a significant performance improvement. It also indicates that INR-based codecs have great potential in compressing multi-view videos, and future research should focus on improving compression performance and coding efficiency.

\vspace{-5pt}
\renewcommand{\baselinestretch}{0.933} 

\bibliography{main}

\begin{thebibliography}{10}

\bibitem{ref1}
D.~Bull and F.~Zhang, {\em Intelligent Image and Video Compression: Communicating Pictures}.
\newblock Academic Press, 2021.

\bibitem{ref2}
G.~Tech, Y.~Chen, K.~Müller, J.-R. Ohm, A.~Vetro, and Y.-K. Wang, ``Overview of the multiview and 3d extensions of high efficiency video coding,'' {\em IEEE Transactions on Circuits and Systems for Video Technology}, vol.~26, no.~1, pp.~35--49, 2016.

\bibitem{ref3}
J.~M. Boyce, R.~Doré, A.~Dziembowski, J.~Fleureau, J.~Jung, B.~Kroon, B.~Salahieh, V.~K.~M. Vadakital, and L.~Yu, ``Mpeg immersive video coding standard,'' {\em Proceedings of the IEEE}, vol.~109, no.~9, pp.~1521--1536, 2021.

\bibitem{ref4}
E.~Dupont, A.~Golinski, M.~Alizadeh, Y.~W. Teh, and A.~Doucet, ``{COIN}: {CO}mpression with implicit neural representations,'' in {\em Neural Compression: From Information Theory to Applications -- Workshop @ ICLR 2021}, 2021.

\bibitem{ref5}
E.~Dupont, H.~Loya, M.~Alizadeh, A.~Golinski, Y.~W. Teh, and A.~Doucet, ``{COIN}++: Neural compression across modalities,'' {\em Transactions on Machine Learning Research}, 2022.

\bibitem{ref6}
Y.~Zhang, T.~van Rozendaal, J.~Brehmer, M.~Nagel, and T.~Cohen, ``Implicit neural video compression,'' in {\em ICLR Workshop on Deep Generative Models for Highly Structured Data}, 2022.

\bibitem{ref7}
H.~Chen, B.~He, H.~Wang, Y.~Ren, S.~N. Lim, and A.~Shrivastava, ``Nerv: Neural representations for videos,'' in {\em Advances in Neural Information Processing Systems}, vol.~34, pp.~21557--21568, Curran Associates, Inc., 2021.

\bibitem{ref8}
Y.~Chen, S.~Liu, and X.~Wang, ``Learning continuous image representation with local implicit image function,'' in {\em Proceedings of the IEEE/CVF Conference on Computer Vision and Pattern Recognition (CVPR)}, pp.~8628--8638, June 2021.

\bibitem{ref9}
A.~Jain, M.~Tancik, and P.~Abbeel, ``Putting nerf on a diet: Semantically consistent few-shot view synthesis,'' 2021.

\bibitem{ref10}
T.~Li, M.~Slavcheva, M.~Zollhoefer, S.~Green, C.~Lassner, C.~Kim, T.~Schmidt, S.~Lovegrove, M.~Goesele, R.~Newcombe, and Z.~Lv, ``Neural 3d video synthesis from multi-view video,'' in {\em 2022 IEEE/CVF Conference on Computer Vision and Pattern Recognition (CVPR)}, pp.~5511--5521, 2022.

\bibitem{ref11}
B.~Mildenhall, P.~P. Srinivasan, M.~Tancik, J.~T. Barron, R.~Ramamoorthi, and R.~Ng, ``Nerf: Representing scenes as neural radiance fields for view synthesis,'' in {\em Proceedings of the European Conference on Computer Vision (ECCV)}, 2020.

\bibitem{ref12}
T.~Müller, A.~Evans, C.~Schied, and A.~Keller, ``Instant neural graphics primitives with a multiresolution hash encoding,'' {\em ACM Transactions on Graphics}, vol.~41, p.~1–15, July 2022.

\bibitem{ref13}
C.~Zhu, G.~Lu, B.~He, R.~Xie, and L.~Song, ``Implicit-explicit integrated representations for multi-view video compression,'' {\em Trans. Img. Proc.}, vol.~34, p.~1106–1118, Jan. 2025.

\bibitem{ref14}
J.~Jung and B.~Kroon, ``Common test conditions for {MPEG} immersive video,'' 2020.

\bibitem{ref16}
G.~Lu, W.~Ouyang, D.~Xu, X.~Zhang, C.~Cai, and Z.~Gao, ``Dvc: An end-to-end deep video compression framework,'' in {\em 2019 IEEE/CVF Conference on Computer Vision and Pattern Recognition (CVPR)}, pp.~10998--11007, 2019.

\bibitem{ref17}
H.~Chen, M.~Gwilliam, S.-N. Lim, and A.~Shrivastava, ``Hnerv: A hybrid neural representation for videos,'' in {\em 2023 IEEE/CVF Conference on Computer Vision and Pattern Recognition (CVPR)}, pp.~10270--10279, 2023.

\bibitem{ref28}
H.~M. Kwan, F.~Zhang, A.~Gower, and D.~Bull, ``Immersive video compression using implicit neural representations,'' in {\em 2024 Picture Coding Symposium (PCS)}, pp.~1--5, 2024.

\bibitem{ref27}
G.~Tech, Y.~Chen, K.~Müller, J.-R. Ohm, A.~Vetro, and Y.-K. Wang, ``Overview of the multiview and 3d extensions of high efficiency video coding,'' {\em IEEE Transactions on Circuits and Systems for Video Technology}, vol.~26, no.~1, pp.~35--49, 2016.

\bibitem{ref19}
T.~Ladune, P.~Philippe, F.~Henry, G.~Clare, and T.~Leguay, ``Cool-chic: Coordinate-based low complexity hierarchical image codec,'' in {\em Proceedings of the IEEE/CVF International Conference on Computer Vision (ICCV)}, pp.~13515--13522, October 2023.

\bibitem{ref20}
H.~Kim, M.~Bauer, L.~Theis, J.~R. Schwarz, and E.~Dupont, ``C3: High-performance and low-complexity neural compression from a single image or video,'' in {\em Proceedings of the IEEE/CVF Conference on Computer Vision and Pattern Recognition (CVPR)}, pp.~9347--9358, June 2024.

\bibitem{ref21}
H.~Chen, M.~Gwilliam, B.~He, S.-N. Lim, and A.~Shrivastava, ``Cnerv: Content-adaptive neural representation for visual data,'' 2022.

\bibitem{ref24}
B.~He, X.~Yang, H.~Wang, Z.~Wu, H.~Chen, S.~Huang, Y.~Ren, S.-N. Lim, and A.~Shrivastava, ``Towards scalable neural representation for diverse videos,'' in {\em 2023 IEEE/CVF Conference on Computer Vision and Pattern Recognition (CVPR)}, pp.~6132--6142, 2023.

\bibitem{ref25}
Z.~Li, M.~Wang, H.~Pi, K.~Xu, J.~Mei, and Y.~Liu, ``E-nerv: Expedite neural video representation with disentangled spatial-temporal context,'' in {\em Computer Vision – ECCV 2022: 17th European Conference, Tel Aviv, Israel, October 23–27, 2022, Proceedings, Part XXXV}, (Berlin, Heidelberg), p.~267–284, Springer-Verlag, 2022.

\bibitem{ref26}
S.~R. Maiya, S.~Girish, M.~Ehrlich, H.~Wang, K.~S. Lee, P.~Poirson, P.~Wu, C.~Wang, and A.~Shrivastava, ``Nirvana: Neural implicit representations of videos with adaptive networks and autoregressive patch-wise modeling,'' in {\em 2023 IEEE/CVF Conference on Computer Vision and Pattern Recognition (CVPR)}, pp.~14378--14387, 2023.

\bibitem{ref37}
G.~Gao, S.~Teng, T.~Peng, F.~Zhang, and D.~Bull, ``Givic: Generative implicit video compression,'' {\em CoRR}, vol.~abs/2503.19604, March 2025.

\bibitem{ref29}
S.~Zhou, Z.~Ni, X.~Zhou, H.~Wen, Y.~Wu, and Y.~Zou, ``Dorefa-net: Training low bitwidth convolutional neural networks with low bitwidth gradients,'' {\em CoRR}, vol.~abs/1606.06160, 2016.

\bibitem{ref30}
D.~A. Huffman, ``A method for the construction of minimum-redundancy codes,'' {\em Proceedings of the IRE}, vol.~40, no.~9, pp.~1098--1101, 1952.

\bibitem{ref36}
G.~Farneb{\"a}ck, ``Two-frame motion estimation based on polynomial expansion,'' in {\em Image Analysis} (J.~Bigun and T.~Gustavsson, eds.), (Berlin, Heidelberg), pp.~363--370, Springer Berlin Heidelberg, 2003.

\bibitem{ref32}
Z.~Wang, A.~Bovik, H.~Sheikh, and E.~Simoncelli, ``Image quality assessment: from error visibility to structural similarity,'' {\em IEEE Transactions on Image Processing}, vol.~13, no.~4, pp.~600--612, 2004.

\bibitem{ref33}
G.~Bjontegaard, ``Calculation of average psnr differences between rd-curves,'' {\em ITU-T VCEG-M33, April, 2001}, 2001.

\end{thebibliography}
\bibliographystyle{ieeetr}

\end{document}